\documentclass[12pt]{article}
\usepackage{amsmath,amsfonts}
\advance \textheight by \topskip
\makeatletter
\@addtoreset{equation}{section}
 
\makeatother
\newcommand{\beq}{\begin{equation}}
\newcommand{\bee}{\end{equation}}
\newcommand{\beqa}{\begin{eqnarray}}
\newcommand{\eeqa}{\end{eqnarray}}

\def\>{\rangle}
\def\<{\langle}

\begin{document}

\title{
%\begin{flushright}
%{\small USACH-FM-01-02}\\[-0.4cm]
%{\small PM-01-07}\\[1cm]
%\end{flushright}
{\bf Four dimensional cubic supersymmetry}}

%Cubic root of translations in field theory }}

\author{
{\sf M.~Rausch de Traubenberg}\thanks{e-mail:
rausch@lpt1.u-strasbg.fr}\\
\\
{\small {\it
Laboratoire de Physique Th\'eorique, CNRS UMR  7085,
Universit\'e Louis Pasteur}}\\
{\small {\it  3 rue de
l'Universit\'e, 67084 Strasbourg, France}}}
\date{}
\maketitle
\vskip-1.5cm

\vspace{2truecm}

\begin{abstract}

\noindent
A four dimensional non-trivial extension of the Poincar\'e algebra
different from  supersymmetry is explicitly studied. Representation
theory is investigated and an invariant Lagrangian is exhibited. 
Some discussion on the Noether theorem is also given.
\end{abstract}

\section{Introduction}

The concept of symmetries is a central tool in the description of 
physical systems. One of the  main questions is, of course, what are the 
mathematical structures  which are useful in describing the laws of physics,
and, in particular, useful in  particle physics or quantum field theories.
Among others, 
finite-dimensional Lie algebras become essential for the description of
space-time symmetries and fundamental interactions.
On the other hand, it was  the discovery of supersymmetry in relativistic 
quantum field theory or as a possible non-trivial 
extension of Poincar\'e invariance
\cite{susy} which gave rise to the concept of Lie superalgebras.
One natural question one should address is the possibility 
to weigh up, in relativistic quantum field theory,
 algebraic structures which are not  Lie (super)algebras.
{\it A priori} this should be a difficult task. Indeed, 
 according to the
Noether theorem, to all  (noetherian)
symmetries correspond  conserved currents. These symmetries are then 
generated by charges which are expressed in terms of the fields.
But, having two kinds of fields, of integer or
half-integer spin, which, by spin-statistics theorem 
 will close with commutators 
or 
anticommutators, {\it a priori}  one should obtain only
Lie and Lie superalgebras.
Furthermore, the
Coleman-Mandula \cite{cm} and the Haag-Lopuszanski-Sohnius \cite{hls} theorems
state that within the framework of Lie algebras one obtains only the
description of space-time  and/or internal symmetries, while within
Lie superalgebras
supersymmetry is the
 {\it unique} non-trivial extension of the Poincar\'e algebra which is 
possible.

But, if one examines the hypotheses of the above theorems, one sees that it
is possible to imagine symmetries which go beyond supersymmetry. Several
possibilities have been considered in the literature,
% \cite{ker, luis,
%fsusy, fsusy1d, fr,am, prs, fsusy2d, fvir, fsusy3d, fsusyh},
the intuitive idea being
that the generators of the Poincar\'e algebra are obtained as an appropriate
product of more fundamental additional symmetries. These new generators
are in an appropriate representation of the Poincar\'e algebra.

In this contribution we would like to study one of the possible
non-trivial extension of the Poincar\'e algebra, different from supersymmetry,
named fractional supersymmetry (FSUSY)\cite{fsusy} -- \cite{cubic}
and its associated underlying algebraic structure named $F-$Lie 
algebras \cite{flie1, flie2}. 
In supersymmetric theories, the extensions of the Poincar\'e
algebra are obtained from a ``square root'' of the translations,
``$QQ \sim P$''.
In this paper,  new algebras
are  obtained from yet higher order roots. We mainly focus on the
simplest alternative  where ``cubic roots'' are involved
``$QQQ \sim P$'' \cite{cubic}. It is important to stress that such structures
are not Lie (super)algebras  
and as such escape {\it a priori} the Coleman-Mandula \cite{cm}
as well as  the Haag-Lopuszanski-Sohnius \cite{hls} no-go theorems.
Furthermore, as far as we know, no no-go theorem associated to such types
of extensions has been considered in the literature. 
This can open interesting  possibilities to search for a field theoretic
realization of a non trivial extension of the Poincar\'e algebra which 
{\it is not the supersymmetric one}.\\ 

The aim of this paper is to summarise some results already obtained
in \cite{cubic} {\it i.e.} to construct explicitly 
 the first field theoretic construction in  
$(1+3)$ dimensions of an FSUSY with $F=3$, which we will 
refer to as {\it cubic supersymmetry}, or  3SUSY. Representation
of our algebra, leads  to fermionic or bosonic multiplets.
 We find that the fermion multiplets are made of three
definite chirality fermions which are degenerate in mass, while the boson 
multiplets contain Lorentz scalars, vectors and two-forms. A striking
feature for the boson multiplets is the compatibility of 3SUSY with gauge
symmetry only when the latter is gauge fixed in the usual way.
Some discussions on Noether theorem in relation with 3SUSY are also given.
The paper is organized as follow.
In the second section  some results on 
the algebraic extension of the Poincar\'e algebra are given.
In section  three representations of the 3SUSY algebra are exhibited.
In section 4, we construct an invariant Lagrangian in the case of the
bosonic multiplet. Section 5 is devoted to some discussions on 
Noether theorem.
Finally, some conclusions are given in section 6. 

\section{Non trivial extension of the Poincar\'e algebra}
 A natural generalization of Lie (super)algebras 
which is relevant for the algebraic description of 
FSUSY was defined in \cite{flie, flie2} and called an $F-$Lie algebra.
An $F-$Lie algebra admits a $\mathbb Z_F-$gradation, the zero-graded part
being a Lie algebra. An $F-$fold symmetric product (playing the role of
the anticommutator in the case $F=2$)  expresses the zero graded  part
in terms of the non-zero graded part. The first examples of $F-$Lie
algebras where infinite-dimensional \cite{flie1}. 
It was then established, that one of these examples of infinite-dimensional 
algebras
was relevant to apply FSUSY on relativistic anyons in $(1+2)D$ \cite{fsusy3d}.
Later on, it was shown how
to construct finite-dimensional $F$-Lie algebras with $F>2$ by an 
inductive process starting from Lie algebras and Lie superalgebras
\cite{flie2}.
Among these families of examples one can identify $F-$Lie algebras that
could generate extensions of the Poincar\'e algebra. One of  these
examples   is given by  

\begin{eqnarray}
\mathfrak{g} = \mathfrak{sp}(4,\mathbb R) \oplus 
\mathrm{ad}\left(\mathfrak{sp}(4,\mathbb R)\right)
\end{eqnarray}

\noindent
with $\mathrm{ad}\left(\mathfrak{sp}(4,\mathbb R)\right)$ the 
adjoint representation of $\mathfrak{sp}(4,\mathbb R)$. 
$\mathfrak{g}_0= \mathfrak{sp}(4,\mathbb R)$ is the zero graded part
(``bosonic'') of $\mathfrak{g}$ and 
$\mathfrak{g}_1=\mathrm{ad}\left(\mathfrak{sp}(4,\mathbb R)\right)$
is the graded part of the algebra.
If we denote
$J_a, \ a=1, \cdots, 10$ a basis of $\mathfrak{sp}(4,\mathbb R)$, 
$A_a$ the corresponding basis for 
$\mathrm{ad}\left(\mathfrak{sp}(4,\mathbb R)\right)$, and 
$g_{ab}= {\mathrm Tr} \left(A_a A_b\right)$ the Killing form of 
$\mathfrak{sp}(4,\mathbb R)$, then  the $F-$Lie algebra of order $3$ 
$\mathfrak{g}$ reads \cite{flie2, cubic}\footnote{In addition to these 
relations, one has also some appropriate Jacobi identities. See
\cite{flie, flie2, cubic} for details.}

\begin{eqnarray}
\label{flie_sp}
\left[J_a, J_b \right] = f_{ab}^{\ \ \ c} J_c, \qquad
 \left[J_a, A_b \right] = f_{ab}^{\ \ \ c} A_c, \qquad
\left\{A_a, A_b, A_c\right\}= g_{ab}J_c + g_{ac} J_b + g_{bc} J_a
\end{eqnarray}

\noindent
where $f_{ab}^{ \ \ c}$ are the structure constant of 
$\mathfrak{sp}(4,\mathbb R)$
and $\left\{A_a, A_b, A_c\right\}$ is given by the symmetric three-fold 
product 
$A_a A_b A_c + A_a A_c A_b + A_b A_c A_a + A_b A_a A_c + A_c A_a A_b +
 A_c A_b A_a$.

\noindent
Observing that $\mathfrak{so}(1,3) \subset \mathfrak{so}(2,3) 
\cong \mathfrak{sp}(4)$, and that the $(1+3)D$ Poincar\'e algebra is related 
to $\mathfrak{sp}(4)$ through an  In\"on\"u-Wigner contraction,  
from the $F-$Lie algebra (\ref{flie_sp}),
 an extension of the Poincar\'e algebra can be constructed \cite{cubic}:
 
\begin{eqnarray}
\label{F-poin2}
&&\left[L_{mn}, L_{pq}\right]=
\eta_{nq} L_{pm}-\eta_{mq} L_{pn} + \eta_{np}L_{mq}-\eta_{mp} L_{nq},
\ \left[L_{mn}, _p \right]= \eta_{np} P_m -\eta_{mp} P_n, \nonumber \\
&&\left[L_{mn}, Q_p \right]= \eta_{np} Q_m -\eta_{mp} Q_n, \ \
\left[P_{m}, Q_n \right]= 0, \\
&&\left\{Q_m, Q_n, Q_r \right \}=
\eta_{m n} P_r +  \eta_{m r} P_n + \eta_{r n} P_m, \nonumber
\end{eqnarray}

\noindent
where $\eta_{m n}$ is  the Minkowski metric, $L_{mn}, P_m$ are the Poincar\'e
generators and $Q_m$ are the ``supercharges'' in the vector representation
of $\mathfrak{so}(1,3)$.

\section{Representations}
Representations of (\ref{F-poin2}) were also studied in \cite{cubic}. 
It turns out that,
for algebras defined by cubic relations the situation is a more 
difficult task than  in usual supersymmetric theories.
Indeed, representations of
supersymmetry are related to representations theory of the well-known
Clifford algebra  while representation theory of FSUSY are  related 
to Clifford algebras of polynomial \cite{cp}.
To obtain  representations of the algebra  (\ref{F-poin2}), we rewrite
the RHS of the trilinear bracket as
$\left\{Q_m, Q_n, Q_r \right \}=f_{mnr} = f_{mnr}^{\ \ \ \ s}P_s$,
with
$f_{mnr}{}^{ s}= \eta_{m n} \delta_r{}^{ s} +  \eta_{m r} \delta_n{}^{ s} +
\eta_{r n} \delta_m{}^{ s}$.  This substitution shows that to the
symmetric tensor $f_{mnr}$ is associated  the cubic polynomial
$f(v^0,v^1,v^2,v^3)= f_{mnr} v^m v^n v^r = 3 (v.P)(v.v)$. Moreover, the
algebra (\ref{F-poin2}) simply  means that $f(v)= \left(v^m Q_m \right)^3$,
as can be verified by  developing the cube and identifying all terms,
using the trilinear bracket.
The generators  $Q_m, m=0, \cdots, 3$ which
are associated with the  variables $v^m,m=0, \cdots, 3$, then generate
an extension of the Clifford algebra
called Clifford algebra of the polynomial $f$
(denoted ${\cal C}_f$). This means that the $Q$'s
allow to ``linearize'' $f$.
The general representation theory  of ${\cal C}_f$ is not known, however 
a systematic method to represent ${\cal C}_f$ by appropriate
 matrices has been given \cite{line}. 
For the algebra (\ref{F-poin2}) one obtains \cite{cubic}

\begin{eqnarray}
\label{Math}
Q_m = \begin{pmatrix}0&\Lambda^{1/3}\gamma_m&0 \cr
                 0&0& \Lambda^{1/3} \gamma_m \cr
                 \Lambda^{-2/3} P_m&0&0 \end{pmatrix}. &
\end{eqnarray}

\noindent
with $\gamma_m$ the  $4D$ Dirac matrices and 
$P_m = - i \frac{\partial~ \ }{\partial x^m}$.
It is interesting to notice that because $P$ is dimensionful, a parameter
with a mass dimension appears naturally in (\ref{Math}). 

The $Q$ being in the vector representation of $\mathfrak{so}(1,3)$, it is
easy to see that 

\begin{eqnarray}
\label{Lorentz}
J_{mn} = \frac14(\gamma_m \gamma_n -\gamma_n \gamma_m) +i(x_m P_n -x_n P_m)
\end{eqnarray}

\noindent   
are the appropriate  Lorentz  generators acting on $Q$: $
%\begin{eqnarray}
 \left[J_{mn},Q_r\right]= \eta_{nr} Q_m - \eta_{mr} Q_m.
%\end{eqnarray}
$

Introducing the $4D$
Dirac matrices in the Weyl representation,
$
\label{eq:gamma}
\gamma_m =
 \begin{pmatrix}
 0&\sigma_m\\
 \bar\sigma_m&0
 \end{pmatrix},
$
with
$
  \sigma_{m\,\alpha\dot\alpha}=\Bigl(1,\sigma_i \Bigr),
 \bar\sigma_m{}^{\dot\alpha\alpha}=
 \Bigl(1,-\sigma_i\Bigr)
$
and $\sigma_i$ the Pauli matrices
(the convention for dotted and undotted indices are those
conventionally used in SUSY, --see {\it e.g.} 
 appendix B of \cite{cubic}--), shows that the representation 
(\ref{Math}) is reducible leading to the two inequivalent $6D$ representations:

\begin{eqnarray}
\begin{array}{ll}
\label{Mat6}
Q_m = \begin{pmatrix}0&\Lambda^{1/3}\sigma_m&0 \cr
                 0&0& \Lambda^{1/3} \bar \sigma_m \cr
                 \Lambda^{-2/3} P_m&0&0 \end{pmatrix},&
Q_m = \begin{pmatrix}0&\Lambda^{1/3}\bar \sigma_m&0 \cr
                 0&0& \Lambda^{1/3}  \sigma_m \cr
                 \Lambda^{-2/3} P_m&0&0 \end{pmatrix}
\end{array}
\end{eqnarray}

\noindent
In this representation the trilinear part of the  
 algebra (\ref{F-poin2}) is realized as

\begin{eqnarray}
\label{susy-ext} 
 Q_m Q_n Q_r + Q_m Q_r Q_n& +& Q_n Q_m Q_r + Q_n Q_r Q_m + Q_r Q_m Q_n +
Q_r Q_n Q_m   \nonumber \\
&=&\eta_{mn} P_r + \eta_{nr} P_m +\eta_{mr} P_n,
 \end{eqnarray}

\subsection{Fermionic multiplet}

As usual, the content of the representation is not only specified by
the form of the matrix representation, but also by the behaviour of the
vacuum under Lorentz transformations. If we denote by $\Omega$ the vacuum,
which is in some specified representation of the Lorentz algebra,
with $\Sigma_{mn}$ the corresponding Lorentz generators, then
$J_{mn}$ given in (\ref{Lorentz}) is  replaced by
$J_{mn} + \Sigma_{mn}$.
In the  case,  where $\Omega$ is a Lorentz scalar,
one sees that the  multiplet of the 
representations (\ref{Mat6}) contains two left-handed   and one
 right-handed fermions for the first matrices, 
while the multiplet of the representation
contains one left-handed  and two right-handed fermions for the second
matrices.
These two multiplets are $CPT$ conjugate.
In the first case, 
if we denote
${\mathbf \Psi}= \begin{pmatrix} \psi_{1 \alpha} \cr \bar \psi_2^{\dot \alpha}
\cr  \psi_{3 \alpha} \end{pmatrix}$, then under a 3SUSY
transformation we have  $\delta_\varepsilon {\mathbf \Psi}
=\varepsilon^m Q_m {\mathbf \Psi}$ and
we obtain 

\begin{eqnarray}
\label{transfo1}
\delta_\varepsilon \psi_{1}{}_\alpha& =& \varepsilon^n
\Lambda^{1/3}\sigma_{n }{}_{\alpha \dot \alpha}
\bar \psi_2{}^{\dot \alpha} \nonumber \\
\delta_\varepsilon \bar \psi_{2}{}^{\dot \alpha} & =& 
\varepsilon^n \Lambda^{1/3}
\bar \sigma_{n}{}^{\dot \alpha  \alpha}
\psi_{3 }{}_\alpha \\
\delta_\varepsilon \psi_{3}{}_\alpha& =& \varepsilon^n
\Lambda^{-2/3}P_n \psi_{1}{}_\alpha \nonumber
\end{eqnarray}

\noindent
with $\varepsilon$ a pure imaginary number.
\subsection{Bosonic multiplet}
In the previous subsection, we were considering the fundamental representation
associated to the matrices (\ref{Mat6}), say
$\mathbf{\Psi}= \begin{pmatrix} \psi_1{}_\alpha \cr
\bar \psi_2{}^{\dot \alpha} \cr \psi_3{}_\alpha \end{pmatrix}$ and
$\mathbf{\Psi^\prime}= \begin{pmatrix}\bar \psi^\prime_1{}^{\dot \alpha} \cr
\psi^\prime_2{}_{ \alpha} \cr \bar \psi^\prime_3{}^{\dot \alpha}
\end{pmatrix}$, {\it i.e.} with the vacuum $\Omega$ in the trivial
representation
of the Lorentz algebra. In this section,  boson multiplets, will
be introduced, corresponding to a vacuum in the spinor representations
of the Lorentz algebra. This means that four types of boson
multiplets can be introduced: $\mathbf{\Psi} \otimes \Omega^\alpha,
\mathbf{\Psi^\prime} \otimes \Omega^\alpha$, with $\Omega^\alpha$ a left-handed
spinor and $\mathbf{\Psi} \otimes \bar \Omega_{\dot \alpha},
\mathbf{\Psi^\prime} \otimes \bar \Omega_{\dot \alpha}$  with
$\bar \Omega_{\dot \alpha}$ a right-handed spinor.

For the multiplet associated to  
$\mathbf{\Psi}^\beta =\mathbf{\Psi} \otimes \Omega^\beta$,  we have
$\mathbf{\Psi}^\beta= \begin{pmatrix} \rho_1{}_{\alpha}{}^{ \beta} \cr 
 \bar \rho_2{}^{\dot \alpha \beta} \cr \rho_3{}_{\alpha }{}^{\beta} 
\end{pmatrix}$ and for the one (CPT conjugate of the previous)
$\mathbf{\Psi}_{\dot \beta} =\mathbf{\Psi^\prime}
\otimes \Omega_{\dot \beta}= \begin{pmatrix} \bar
\rho_1{}^{\dot\alpha}{}_{ \dot \beta} \cr
  \rho_2{}_{ \alpha \dot  \beta} \cr \bar
\rho_3{}^{\dot \alpha }{}_{\dot \beta}
\end{pmatrix}$.
The transformation  under 3SUSY is 
$\delta_\varepsilon \mathbf{\Psi}^\beta = \varepsilon^m Q_m  
\mathbf{\Psi}^\beta$
with $Q_m$ given in (\ref{Mat6}) and similarly for
$\mathbf{\Psi}_{\dot \beta}$. %This leads to

%\begin{eqnarray}
%\label{transfo2}
%\delta_\varepsilon \rho_1{}_{\alpha}{}^{ \beta}& =& \Lambda^{1/3}
%\varepsilon^m \sigma_m{}_{\alpha \dot \alpha}
%\bar \rho_2{}^{\dot \alpha \beta} \nonumber \\
%\delta_\varepsilon \bar \rho_2{}^{\dot \alpha \beta} &=&\Lambda^{1/3}
%\varepsilon^m \bar \sigma_m{}^{\dot \alpha  \alpha} 
%\rho_3{}_{\alpha}{}^{ \beta}
%\\
%\delta_\varepsilon \rho_3{}_{\alpha}{}^{ \beta}& =& \Lambda^{-2/3} 
%\varepsilon^m P_m
%\rho_1{}_{\alpha}{}^{ \beta} . \nonumber 
%\end{eqnarray}

\noindent
Notice that $\rho_1, \bar \rho_1,  \bar \rho_2,\rho_2$ and 
$\rho_3, \bar \rho_3$ are  not irreducible 
representations
of $\mathfrak{sl}(2,\mathbb C) \cong \mathfrak{so}(1,3)$, we therefore
define 

\begin{eqnarray}
\label{bos-fields}
\begin{array}{ll}
\rho_1= \varphi \,I_2 + \frac{1}{2} B_{mn} \,\sigma^{nm} &
\bar \rho_1= \varphi^\prime \,\bar I_2 + \frac{1}{2} B^\prime_{mn}
\bar \sigma^{nm}
\cr
\bar \rho_2= A^m \, \bar \sigma_m &
\rho_2= A^\prime{}^m \,  \sigma_m 
\cr
\rho_3= \tilde \varphi \, I_2 + \frac{1}{2} \tilde B_{mn}\, \sigma^{nm}&
\bar \rho_3= \tilde \varphi^\prime \, \bar I_2 + \frac{1}{2} \tilde 
B^\prime{}_{mn}\,
\bar \sigma^{nm}
\end{array}
\end{eqnarray}

\noindent
with $I_2$ and $\bar I_2$ the two by two identities matrix,
$\sigma_{mn}$ and $\bar \sigma_{mn}$ the Lorentz generators for
the two spin representations,
$A^m$ and $A^\prime{}^m$ two vectors, $\varphi, \tilde \varphi$
and   $\varphi^\prime , \tilde \varphi^\prime$ four scalars, 
$B_{mn},\tilde B_{mn}$ two self-dual two-forms
and 
$B^\prime{}_{mn},\tilde B^\prime{}_{mn}$ two anti-self-dual two-forms.
Then one can show  that  the transformations reads  \cite{cubic}

\begin{eqnarray}
\label{transfo2vect}
\begin{array}{ll}
\delta_\varepsilon \varphi =\Lambda^{1/3}  \varepsilon^m A_m &
\delta_\varepsilon \varphi^\prime =  \Lambda^{1/3}
\varepsilon^m A^\prime_m
\cr
\delta_\varepsilon B_{mn} = - \Lambda^{1/3}\left(
\varepsilon_m A_n - \varepsilon_n A_m \right)  &
%+\Lambda^{1/3} i  
%\varepsilon_{mnpq} \varepsilon^p A^q  &
\delta_\varepsilon B^\prime_{mn} = - \Lambda^{1/3}\left(
\varepsilon_m A^\prime_n - \varepsilon_n A^\prime_m \right) 
%-\Lambda^{1/3} i  
%\varepsilon_{mnpq} \varepsilon^p A^\prime{}^q
\cr
\ \ \ \ \ \ \ \ \ +\Lambda^{1/3} i  
\varepsilon_{mnpq} \varepsilon^p A^q  &
\ \ \ \ \ \ \ \ \    -\Lambda^{1/3} i  
\varepsilon_{mnpq} \varepsilon^p A^\prime{}^q
\cr
\delta_\varepsilon A_m = \Lambda^{1/3}\left( \varepsilon_m \tilde \varphi +
 \varepsilon^n \tilde B_{mn} \right) &  
\delta_\varepsilon A^\prime_m =  \Lambda^{1/3}\left(
\varepsilon_m \tilde \varphi^\prime + 
\varepsilon^n \tilde B^\prime_{mn}\right) 
\cr
\delta_\varepsilon \tilde \varphi = \Lambda^{-2/3} 
\varepsilon^m P_m \varphi &
\delta_\varepsilon \tilde \varphi^\prime = \Lambda^{-2/3}\varepsilon^m 
P_m \varphi^\prime
\cr
\delta_\varepsilon  \tilde B_{mn} = \Lambda^{-2/3}\varepsilon^p P_p B_{mn}&
\delta_\varepsilon  \tilde B^\prime_{mn} = \Lambda^{-2/3}\varepsilon^p P_p 
B^\prime_{mn}
\end{array}
\end{eqnarray}

The second bosonic multiplet being
CPT conjugate to the first one, we have
$\left(\bar \rho_1{}^{\dot \alpha}{}_{\dot \beta}\right)^\star=
\rho_1{}^{\alpha}{}_{\beta},
\left(\rho_{2 \alpha \dot \beta}\right)^\star=
\bar \rho_{2 \dot \alpha  \beta}$ and
$\left(\bar \rho_3{}^{\dot \alpha}{}_{\dot \beta}\right)^\star=
\rho_3{}^{\alpha}{}_{\beta}$. 
[$B^\star$, the complex conjugate of $B$,
is  not to be confused with ${}^\star B$, the dual of $B$, see after.]
This means, paying attention to the position
of the indices  that we have

\begin{eqnarray}
\label{cc}
\begin{array}{ll}
\varphi^\prime{}^\star = - \varphi,&
 \tilde \varphi^\prime{}^\star = -\tilde \varphi, \cr
B_{mn}^{\prime \, \star}= -B_{mn},&
\tilde B_{mn}^{\prime \, \star}= -\tilde B_{mn}, \cr
A^{\prime \, \star}_m= A_m.
\end{array}
\end{eqnarray}

\noindent
These relations are compatible with the transformations laws given
in (\ref{transfo2vect})  since
$\varepsilon_n{}^\star = - \varepsilon_n$ and $P_n= -i \frac{\partial}{\partial
x^n}$.

\section{Invariant action}
To construct a real invariant action involving the bosonic multiplets
(\ref{bos-fields}), we introduce the real fields

\begin{eqnarray}
\label{real}
A_-=i\frac{A-A^\prime}{\sqrt{2}} &,&
A_+= \frac{A+A^\prime}{\sqrt{2}}, \nonumber \\
B_-= \frac{B-B^\prime}{\sqrt{2}} &,& 
B_+= i\frac{B+B^\prime}{\sqrt{2}}, \nonumber \\
\tilde B_-=  \frac{\tilde B-\tilde B^\prime}{\sqrt{2}} &,& 
\tilde B_+=  i\frac{\tilde B+\tilde B^\prime}{\sqrt{2}}, \\
\varphi_-= \frac{\varphi-\varphi^\prime}{\sqrt{2}} &,& 
\varphi_+= i\frac{\varphi+\varphi^\prime}{\sqrt{2}}, \nonumber \\
\tilde \varphi_-=  \frac{\tilde \varphi-\tilde \varphi^\prime}{\sqrt{2}} &,& 
\tilde \varphi_+=  i\frac{\tilde \varphi+\tilde \varphi^\prime}{\sqrt{2}}. 
\nonumber
\end{eqnarray} 

\noindent
These new fields form now one (reducible) multiplet of 3SUSY,  
 with ${}^\star B_- =  B_+$ (${}^\star B_-$ is the dual of $B_-$ \cite{cubic}).
The corresponding  two- and three-form field strengths read

\begin{eqnarray}
\label{curl}
&&F_{\pm}{}_{mn}=\partial_m A_{\pm}{}_n -\partial_n A_{\pm}{}_m, \  
%F^\prime _{mn}=\partial_m A^\prime_n -\partial_n A^\prime_m, 
\nonumber \\
&&H_{\pm}{}_{mnp}= \partial_m B_{\pm}{}_{np} + \partial_{n} 
B_{\pm}{}_{pm} + 
\partial_p B_{\pm}{}_{mn}.
\end{eqnarray} 

\noindent
They  are  invariant under the gauge transformations

\begin{eqnarray}
\label{form}
\varphi_\pm &\to & \varphi_\pm   \nonumber \\
A_{\pm}{}_m &\to & A_{\pm}{}_m + \partial_m \chi_{\pm} \\ 
B_{\pm}{}_{mn} &\to & B_{\pm}{}_{mn} + 
(\partial_m \chi_{\pm \,n} -\partial_n \chi_{\pm \, _m} ) \nonumber
\end{eqnarray}

\noindent
where $\chi_{\pm}$ ($\chi_{\pm}^m$) are arbitrary scalar (vector)
functions  ($ \chi^m_{-}$ and    $\chi^m_+$ can still be related in order
to preserve the duality relations between $B_-$ and $B_+$)\footnote{
Note that the gauge transformations (\ref{form}) correspond naturally
to the zero-, one- and two-form character of the components of the 3SUSY
gauge multiplet.}.
 
In a similar way we introduce the field strength
$\tilde H_-{}_{mnp}, \tilde H_+{}_{mnp}$, as well as the dual fields 
${}^\star H_-{}_m,
{}^\star H_+{}_m,
{}^\star\tilde  H_-{}_m, {}^\star \tilde H_+{}_m$
(where ${}^\star H_m \equiv \frac{1}{6} \varepsilon_{mnpq} H{}^{npq}$.
For instance ${}^\star \tilde H_-{}_m = i \partial^n B_+{}_{m n}$).
We consider now the following local gauge invariant and zero graded 
Lagrangian,

\begin{eqnarray}
\label{Lbos}
{\cal L}& = &\partial_m \varphi_- \partial^m \tilde \varphi_-  -
           \partial_m \varphi_+  \partial^m \tilde \varphi_+  \nonumber \\
&-&\frac{1}{4} F_{-}{}_{mn}F_-{}^{mn} +\frac{1}{4} F_{+}{}_{mn}F_+{}^{mn} 
-\frac12\left(\partial_m A_-{}^m\right)^2 + 
\frac12\left(\partial_m A_+{}^m\right)^2 
\\
&-& \frac{1}{12} H_-{}_{mnp} \tilde H_-{}^{mnp} 
+\frac{1}{12} H_+{}_{mnp} \tilde H_+{}^{mnp}+
\frac{1}{2} {}^\star H_-{}_m  {}^\star \tilde H_-{}^{m}-
\frac{1}{2} {}^\star H_+{}_m  {}^\star \tilde H_+{}^{m}
 \nonumber
\end{eqnarray}

\noindent
Using (\ref{real}), a  direct calculation shows that (\ref{Lbos}) is 
invariant under the 
transformations (\ref{transfo2vect}), up to a surface term.
It is interesting to notice that the 3SUSY invariance is compatible with
gauge symmetries if the latter are gauged fixed.
A usual 't Hooft Feynman gauge fixing term 
($-\frac12\left(\partial_m A_-{}^m\right)^2
   + \frac12\left(\partial_m A_+{}^m\right)^2$) is required for 
the vector fields.
For the two forms, due to the relation
 ${}^\star B_-= B_+$ also some gauge fixing terms
{\it \`a la}  't~Hooft Feynman
are   present. Developing all terms in (\ref{Lbos})
the Lagrangian  can be rewritten {\it \`a la }  ``Fermi-like''

\begin{eqnarray}
\label{Lbos2}
{\cal L}& = &\partial_m \varphi_- \partial^m \tilde \varphi_-  
-\frac{1}{2} \partial_m A_-{}_n \partial^m A_-{}^n 
+ \frac{1}{4} \partial_m B_-{}_{np}  \partial^m \tilde B_-{}_{np}  \nonumber \\
&-&\partial_m \varphi_+  \partial^m \tilde \varphi_+  
+\frac{1}{2} \partial_m A_+{}_n \partial^m A_+{}^n 
-  \frac{1}{4} \partial_m B_+{}_{np}  \partial^m \tilde B_+{}_{np} 
 \nonumber
\end{eqnarray}

\noindent
(similar Lagrangian appear in the action-at-a-distance formalism for
vector and two-forms --see {\it e.g} \cite{rk}).

One should note, though, the relative minus signs in front of the kinetic
terms of the vector fields in (\ref{Lbos}) which endanger {\sl a priori} 
the boundedness from below of the density  energy of the 
``electromagnetic" fields. This difficulty does not have a clear physical
interpretation as long as interaction terms have not been included, and 
necessitates a more careful study of the field manifold
associated to the density energy.

\section{ Noether currents}
The 3SUSY algebra we have studied  has one main difference 
with the usual Lie (super)algebra: it does not close through quadratic,
but rather cubic, relations. Moreover, it might be possible that
some usual results of Lie (super)algebra do not apply straightforwardly.
One example is the  Noether currents and their associated algebra.
This interesting point was studied in \cite{cubic}. 

In this section we would like, however, to construct explicitly the
Noether current. Using (\ref{Lbos}) and (\ref{transfo2vect}) we obtain

\begin{eqnarray}
J_{m n}& =& -i  \Lambda^{1/3} ( A_-{}_n  \partial_m
 \tilde \varphi_-
- \partial_m   A_-{}_n  \tilde \varphi_- )  
 -i  \Lambda^{1/3} ( A_+{}_n  \partial_m
  \tilde \varphi_+
 -  \partial_m   A_+{}_n  \tilde \varphi_+ )  \nonumber \\
&+&i  \Lambda^{1/3}(\tilde B_-{}_{r n } \partial_m
 A_-^r
 -  \tilde B_-{}_{r n }   \partial_m A_-^r )
+i  \Lambda^{1/3}( \tilde B_+{}_{r n } \partial_m
 A_+^r
- \tilde B_+{}_{r n }    \partial_m A_+^r )
\\
&-&i  \Lambda^{-2/3}( \partial_m   \varphi_- 
\partial_n   \varphi_-
 - \partial_m  \varphi_+  \partial_n   \varphi_+ )
-\frac{i}{4}  \Lambda^{-2/3} ( \partial_m  B_-^{r s}
 \partial_n  B_{- r s}  -
 \partial_m  B_+^{r s}
 \partial_n  B_{+ r s}) \nonumber
\end{eqnarray}

\noindent
which is, due to the equations of motion, 
conserved $\partial_m J^{mn}=0$ (up to a surface term).
Then, the conserved charges are obtained as usual
$\hat Q_m = \int d^3 x J_{0m}
$.
Introducing the conjugate momentum $\Pi$ 
of the fields $\Psi$ (where $\Psi$ is one of the fields of section 4)  :
$\Pi = \frac{\delta {\cal L}}
{\delta \partial_0 \Psi}$ after the usual quantization (equal-time commutation
relations) one easily obtains 

\begin{eqnarray}
\label{quantum}
\delta_\varepsilon { \Psi} = \left[\varepsilon_n \hat Q^n, \Psi\right]
\ \ \ 
\Big(\delta_{Q_m}{ \Psi} = \left[ \hat Q^m, f{ \Psi}\right]\Big)
\end{eqnarray}

\noindent
In particular this means that the algebra (\ref{F-poin2}) is realized 
through

\vskip -,5truecm 
\begin{eqnarray}
\label{F-com}
\begin{array}{lll}
\left( \delta_{Q_m}. \delta_{Q_n}.
\delta_{Q_r} + \mathrm{perm} \right) { \Psi} &=&  \cr
\left[\hat Q_m, \left[\hat Q_n, \left[\hat Q_r,  
{ \Psi} \right]\right]\right]+ \mathrm{perm} &=&
\eta_{m n} \left[\hat P_r,{ \Psi} \right] +  
\eta_{m r} \left[\hat P_n,{ \Psi} \right] + 
\eta_{r n} \left[\hat P_m, { \Psi} \right] \cr
&=&\left(\eta_{m n} \delta_{P_r} + \eta_{m r} \delta_{P_n} +
\eta_{r n} \delta_{P_m}\right){ \Psi}
\end{array}
\end{eqnarray}

\noindent
with $\hat P$ the generators of the Poincar\'e translations.
Indeed, starting with the abstract algebra (\ref{F-poin2}), we
can represent it by some matrices as in section 3 (see {\it e.g.}
(\ref{Math})). In this case the product of two transformations will be given
by $\delta_n \delta_m \Psi = Q_n Q_m \Psi$
and the algebra will be realized as in  (\ref{susy-ext}). 
But, we can also represent  (\ref{F-poin2}) with commutators (\ref{quantum})
acting on some Hilbert space, thus 
the product of two transformations will be given by 
$\delta_n \delta_m \Psi = \left[\hat Q_n, \left[\hat Q_m, \Psi \right]\right]$
leading to
the realization (\ref{F-com}) of the algebra (\ref{F-poin2}).
For a more general discussion one can see \cite{cubic}.

\section{Conclusion}
In this paper we have  studied some four dimensional realizations
of 3SUSY, a non-trivial extension of the Poincar\'e algebra
different from supersymmetry. Representation
theory was explicitly constructed. Then, an invariant Lagrangian 
involving bosonic fields was given (some Lagrangian involving
fermionic fields was also considered in \cite{cubic}). 
The next step, will be to construct an interacting theory.

%\begin{thebibliography}{99}

%\end{thebibliography}\LastPageEnding

\end{document}